\documentclass[twocolumn,showpacs,preprintnumbers,amsmath,amssymb]{revtex4}

\usepackage{graphicx}
\usepackage{dcolumn}
\usepackage{bm}
\usepackage{paralist}
\usepackage{amsmath}
\usepackage{psfrag}
\usepackage{subfig}
\usepackage{color}
\newcommand{\avg}[1]{\overline{#1}}
\newcommand{\spavg}[1]{\left\langle{#1}\right\rangle_{(t)}}
\newcommand{\vex}[1]{\mathbf{#1}}
\newcommand{\ten}[1]{\mathbf{#1}}
\newcommand{\tentau}{\mathbf{T}}
\newcommand{\vnabla}{\boldsymbol{\nabla}}

\newcommand{\pdert}[1]{\frac{\partial #1}{\partial t}}

\newcommand{\eqnref}[1]{Eq.~(\ref{#1})}
\newcommand{\figref}[1]{Fig.~\ref{#1}}

\def\square{${\vcenter{\hrule height .8pt
        \hbox{\vrule width .8pt height 5pt \kern 5pt
        \vrule width .8pt}
        \hrule height .8pt}}$}

\newcount\ndots
\def\drawline#1#2{\raise 2.5pt\vbox{\hrule width #1pt height #2pt}}
\def\spacce#1{\hskip #1pt}
\def\solid{\drawline{24}{.5}\nobreak\ }
\def\boldsolid{\drawline{24}{2.}\nobreak\ }
\def\bdash{\hbox{\drawline{4}{.5}\spacce{2}}}
\def\dashed{\bdash\bdash\bdash\bdash\nobreak\ }

\def\bdot{\hbox{\drawline{1}{.5}\spacce{2}}}
\def\dotted{\hbox{\leaders\bdot\hskip 24pt}\nobreak\ }

\def\trian{\raise 1.25pt\hbox{$\scriptscriptstyle\triangle$}\nobreak\ }
\def\circle{$\circ$\nobreak\ }

\def\square{${\vcenter{\hrule height .4pt
        \hbox{\vrule width .4pt height 3pt \kern 3pt
        \vrule width .4pt}
        \hrule height .4pt}}$\nobreak\ }
\def\plus{\raise 1.25pt \hbox{$\scriptscriptstyle +$}\nobreak\ }
\def\x{\raise 1.25pt \hbox{$\scriptscriptstyle \times$}\nobreak\ }

\def\solidtrian{\raise 1.25pt
   \hbox to 3bp{
\hfill}\nobreak\ }
\def\solidsquare{\vrule height .9ex width .8ex depth -.1ex\nobreak\ }

\def\solidcclose{\drawline{10}{.5}\nobreak\raise
  0.5pt\hbox{$\bullet$}\drawline{10}{.5}\nobreak\ }

\def\solidsclose{\drawline{10}{.5}\nobreak\raise
  0.5pt\hbox{\solidsquare}\drawline{10}{.5}\nobreak\ }

\def\solidtclose{\drawline{10}{.5}\nobreak\raise
  0.5pt\hbox{\solidtrian}\drawline{10}{.5}\nobreak\ }

\def\solidcopen{\drawline{10}{.5}\nobreak\raise
  0.5pt\hbox{\circle}\drawline{10}{.5}\nobreak\ }

\def\solidsopen{\drawline{10}{.5}\nobreak\raise
  0.5pt\hbox{\square}\drawline{10}{.5}\nobreak\ }

\def\solidtopen{\drawline{10}{.5}\nobreak\raise
  0.5pt\hbox{\trian}\drawline{10}{.5}\nobreak\ }

\def\solidx{\drawline{10}{.5}\nobreak\raise
  0.5pt\hbox{\x}\drawline{10}{.5}\nobreak\ }

\begin{document}

\preprint{APS/123-QED}

\title{Polymer Maximum Drag Reduction: A Unique Transitional State
}

\author{Yves Dubief$^1$, C. M. White$^2$, E. S. G. Shaqfeh$^3$ and V. E. Terrapon$^4$}
\affiliation{%
$^1$School of Engineering,
 University of Vermont,
 Burlington VT\\
$^2$Department of Mechanical Engineering,
 University of New Hampshire,
 Durham NH\\
$^3$Department of Mechanical Engineering,
 Stanford University,
 Stanford CA \\
$^4$Aerospace and Mechanical Engineering Department,
 University of Liege,
 Belgium
}%

\date{\today}

\begin{abstract}

The upper bound of polymer drag reduction is identified as a unique
transitional state between laminar and turbulent flow corresponding
to the onset of the nonlinear breakdown of flow instabilities.

\end{abstract}

\pacs{Valid PACS appear here}
\maketitle


Drag reduction by the addition of long-chain polymers to flowing
liquids is bounded by the so-called maximum drag reduction (MDR)
asymptote \cite{virk1970uaa}. Drag reductions attained at MDR may be
as high as 80\% in turbulent boundary layers and channel flows over
smooth plane walls, and as such, MDR has received extensive research interest. One characteristic of MDR is that the mean
velocity profile, determined empirically by Virk \textit{et
al.}\cite{virk1970uaa} as:
\begin{equation}
\avg{u}^+=11.7\log(y^+)-17, \label{eq:virk}
\end{equation}
is found to be roughly universal, insensitive to polymer species,
molecular weight, or the polymer-solvent pair. The superscript $^+$
in \eqnref{eq:virk} denotes normalization by the friction velocity
$u_\tau = \sqrt{\tau_w / \rho}$ and kinematic viscosity $\nu$, where
$\tau_w$ is the shear stress at the wall and $\rho$ the fluid
density. While the universality of the Virk profile is generally
well-accepted, a point of controversy is the behavior of the
Reynolds shear stress at MDR. Warholic \textit{et
al.}\cite{warholic1999influence} measured virtually no Reynolds
shear stress and conjectured that MDR corresponds to the state where
Reynolds stresses are negligible and turbulence is sustained by
polymer stresses. However, Ptasinski \textit{et al.}
\cite{ptasinski2001etp} reports a reduced, yet finite Reynolds shear
stress at MDR, corresponding to approximately a 50\% reduction in
maximum Reynolds shear stress compared to a Newtonian flow.
Nevertheless, despite the differences in high-order turbulence statistics, both
experiments find a mean velocity profile that agrees reasonably well
with Eq. (\ref{eq:virk}). The question therefore arises whether MDR
is uniquely defined and, as such, can it be uniquely predicted. The
answer to this question is important not only to better understand
the bounding mechanisms of MDR but also to develop robust models to
predict drag reduction given an initial set of conditions.

\begin{figure*}
\begin{center}
\vspace*{0.5cm}
\begin{minipage}{0.32\textwidth}
\subfloat{\label{fig:Wu-U+}
\psfrag{(a)}{(a)}
\psfrag{U+}{$\avg{u}^+$}
\psfrag{y+}{$y^+$}
\includegraphics[width=1\textwidth]{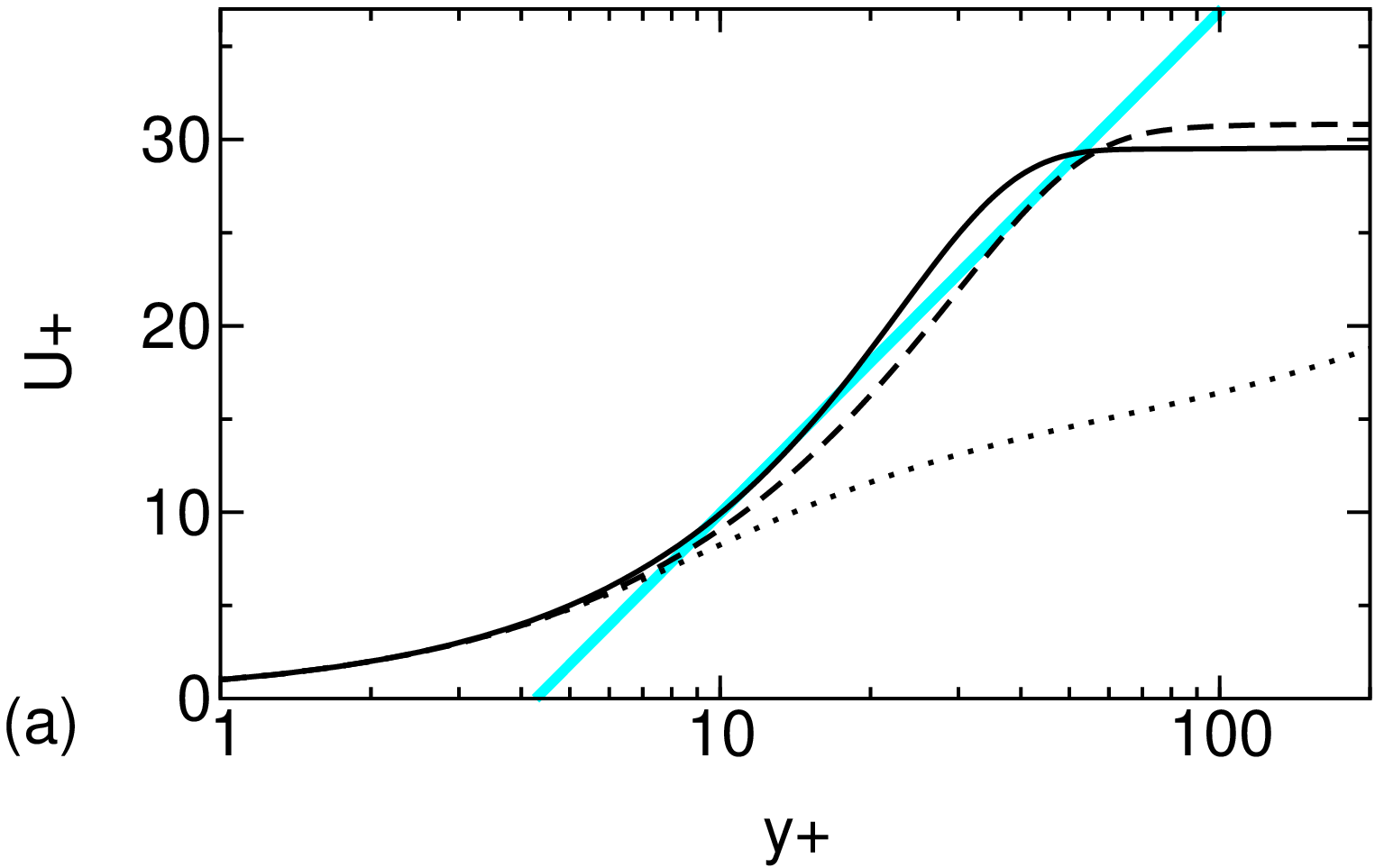}}
\end{minipage}
\hfill
\begin{minipage}{0.32\textwidth}
\subfloat{\label{fig:Wu--uv+}
\psfrag{(a)}{(b)}
\psfrag{-uv+}{$-\avg{uv}^+$}
\psfrag{y+}{$y^+$}
\includegraphics[width=1\textwidth]{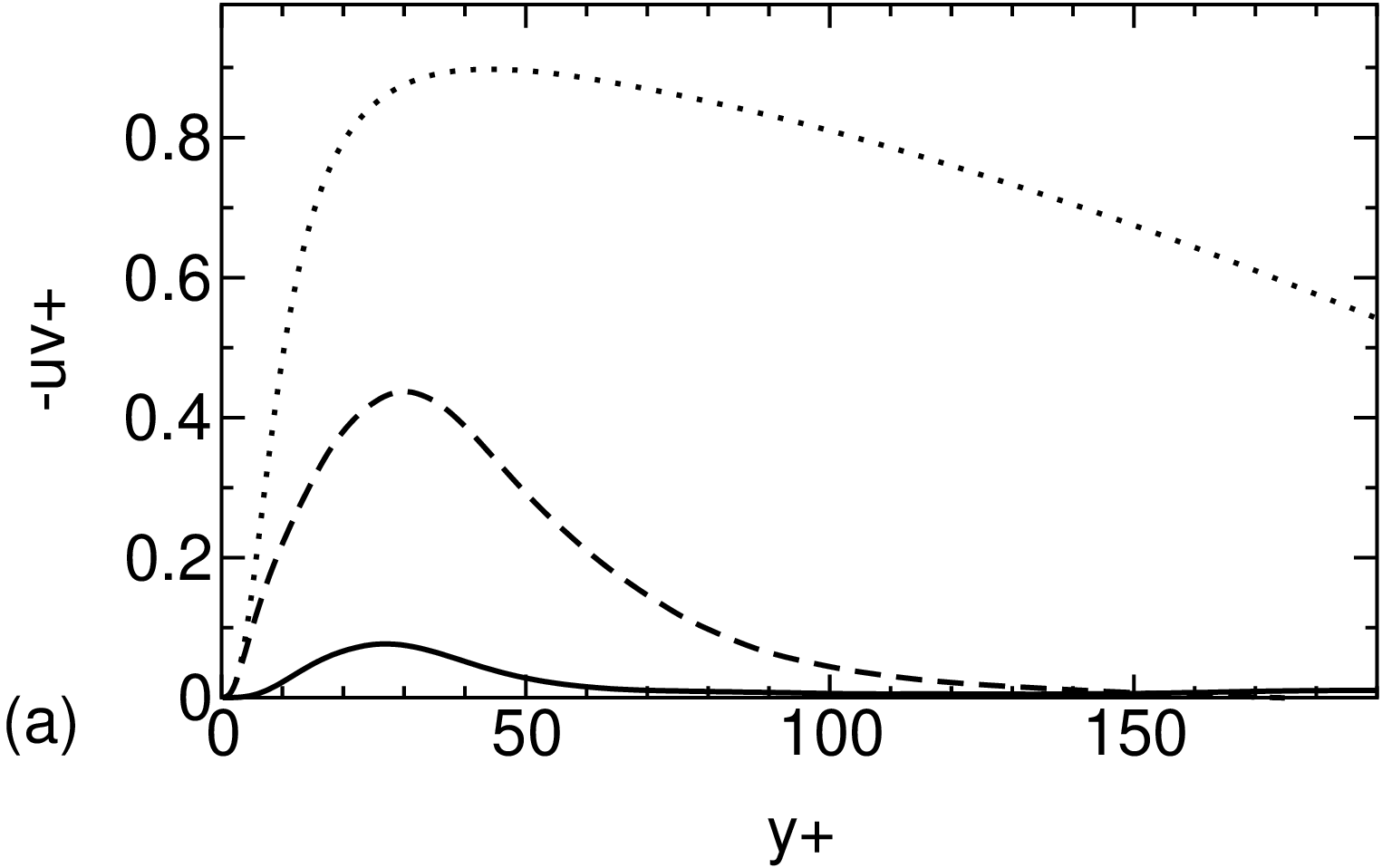}}
\end{minipage}
\hfill
\begin{minipage}{0.32\textwidth}
\subfloat{\label{fig:Wu-Cf}
\psfrag{(a)}{(c)}
\psfrag{Cf.103}{$C_f\times10^3$}
\psfrag{rey}{$Re_\theta$}
\includegraphics[width=1\textwidth]{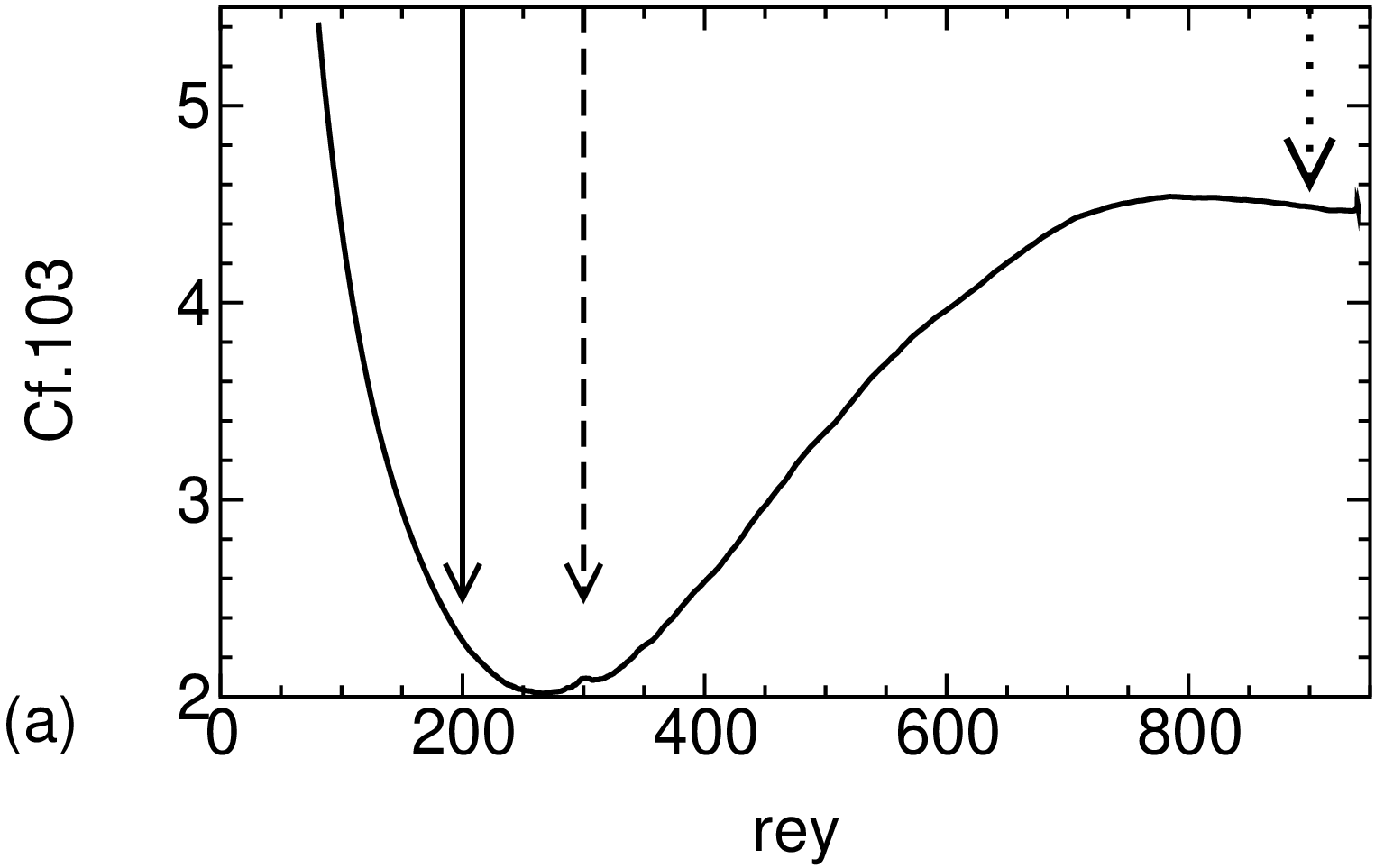}}
\end{minipage}
\end{center}
\caption{\label{fig:Wu-MDR} Velocity statistics of the transitional boundary layer simulation of \cite{wu2009direct} for two states whose velocity profiles approach closely Virk's asymptotic velocity profile (Eq.~\ref{eq:virk}). (a) Shows \eqnref{eq:virk} \color{cyan}$\boldsolid$\color{black}, $Re_\theta=200$ \solid,  $Re_\theta=300$ \dashed, and  $Re_\theta=900$ \dotted (fully turbulent flow). (b) Compares Reynolds shear stress, $-\avg{uv}^+$, for the three flow states as shown in (a). (c) Plots the distribution of the skin friction coefficient $C_f$ as a function of the Reynolds number based on the local momentum thickness $\theta$. Arrows point to the statistical profiles plotted in (a) and (b).
}
\end{figure*}

This letter tests and demonstrates the hypothesis that MDR is a
unique transitional state between laminar and turbulent flow and
that the state of turbulence at MDR can be extrapolated by the
mechanisms of polymer-vortex interactions framed by Dubief \text{et
al.}\cite{dubief2004cdr}, Terrapon \textit{et
al.}\cite{terrapon2004sps} and Kim \text{et
al.}\cite{kim2007effects}. A necessary condition of our hypothesis is the existence of transitional states in
a Newtonian fluid with velocity statistics similar to MDR states
observed experimentally \cite{warholic1999influence,
ptasinski2001etp}. To demonstrate that this condition is satisfied, we have analyzed the DNS  data of bypass
transition in a boundary layer flow of Wu \&
Moin\cite{wu2009direct}. \figref{fig:Wu-U+} indicates that the
region between $Re_\theta=200$ and 300 best approaches Virk's
asymptotic velocity profile. Here the Reynolds number, $Re_\theta$,
is based on the momentum boundary layer thickness $\theta$. Shown in
\figref{fig:Wu--uv+} is that from $Re_\theta=200$ to
$Re_\theta=300$, the Reynolds shear stress increases from almost
negligible levels ($\lesssim10\%$) of the full turbulent state to
roughly half the fully turbulent level. In between these two states,
the skin friction $C_f$ (\figref{fig:Wu-Cf}) reaches a minimum,
which marks the end of the instability stage and the beginning of
the nonlinear development stage. Collectively, these data suggest
that the MDR state may correspond to a narrow region near the
beginning of the nonlinear breakdown stage, including pre- and
post-breakdown stages. They are also suggestive that the variations
in high-order turbulence statistics, such as Reynolds shear stress, observed
experimentally at MDR may likely be a consequence of  the transitioning between  pre- and post-breakdown stages when computing averaged turbulence statistics. 

Viewed from the framework of the underlying fluid structure, the
nonlinear breakdown stage is associated with the breakdown of the
near-wall low- and high-speed velocity streaks and the subsequent
generation of near-wall quasi-streamwise vortices. Once triggered,
the wall-shear, velocity streaks, and quasi-streamwise vortices
form the autonomous self-sustaining cycle of wall turbulence
\cite{jimenez1999autonomous}. The low- and high-speed streaks span
from the viscous sublayer, a thin region close to the wall where
viscous dissipation dominates, to the adjacent so-called buffer
layer. The quasi-streamwise vortices evolve in the buffer layer,
around the streaks, and create upwash and downwash flows that (a) are biaxial extensional flows\cite{terrapon2004sps} and (b)
contribute to the vertical transfer of momentum throughout the
viscous, buffer and log layers.

The effect of polymers on turbulent wall-flow is to disrupt and
modify the near-wall autonomous cycle of wall turbulence that
results in a net drag reduction. Previous studies investigating
polymer-vortex interactions \cite{dubief2004cdr,
terrapon2004sps,kim2007effects} demonstrated that polymers primarily
stretch (and extract energy) in the upwash and downwash regions
generated by quasi-streamwise vortices and, consequently, dampen the
vortices. With increasing drag reduction, vortices become
increasingly weaker and the high- and low-speed streaks become more
coherent \cite{white2004tsd}.

To best frame our hypothesis, let us first consider two thought experiments
using polymers that possess the necessary elasticity to eradicate
vortices. In the first experiment, consider that this polymer
additive is suddenly added to fully-developed Newtonian turbulence (say by some
injection scheme). Based on our understanding of polymer-turbulence
interactions, it is expected that after an initial transient time
in which coiled polymers become sufficiently stretched in the
biaxial-extensional flows around vortices, the vortices will
disappear but streaks will remain. Once all vortices are damped out, the
stretching mechanism for polymers is limited to wall-shear and
spanwise shear layers between streaks, both modest source of polymer
stretching compared to biaxial-extensional flow
\cite{terrapon2004sps}. Polymers therefore recoil, allowing
instabilities to grow and new vortices to form. The emerging
vortices grow until their biaxial-extensional flows become
sufficiently strong to stretch the coiled polymers and the cycle
repeats itself.

For the second thought experiment, let us consider a transitional
flow of a polymer solution (i.e., a polymer ocean), again with the
necessary elasticity to eradicate vortices. Since the
turbulence-reducing mechanisms of polymers requires vortices to be
effective, the impact of polymers on the linear region of the
transition is expected to be marginal (note that polymers have no
drag reducing effect on wall-bounded laminar flows). However, at the
onset of the nonlinear breakdown of flow instabilities and the
formation of vortices, the polymer drag reducing mechanisms are
activated, the vortices become damped, and the flow is returned to a
pre-breakdown transitional stage. This scenario is very much
consistent with experimental results that show that polymer
solutions that transition from laminar to MDR, stay at MDR as the
Reynolds number is increased \cite{virk1975drag}.

Based on the two thought experiments described above,  knowledge
of the mechanisms of polymer-turbulence interactions, and the demonstrated existence of transitional states in a Newtonian fluid with velocity statistics similar to MDR, the hypothesis presented here is that a precise definition of MDR is that it is the transitional state that  corresponds to the onset of the nonlinear
breakdown stage of transition. In practice, however, MDR turbulence corresponds to a narrow flow region centered about the onset of the nonlinear breakdown stage of transition that includes pre- and post-breakdown stages. We provide further substantiative
evidence that this hypothesis is correct using direct numerical
simulation of transition in a channel flow with polymer additives.
The motivation to use a channel flow over a boundary layer
simulation stems from the need to demonstrate a sustained MDR state.
In a boundary layer, the required streamwise length scale of the domain for accurate computation and flow development is prohibitively large, whereas,  a channel flow may be reasonably calculated over a very long time series.

\begin{figure*}[t]
  \begin{center}
\begin{minipage}{0.32\textwidth}
\subfloat{\label{fig:trans-post-U+}
\psfrag{(a)}{(a)}
\psfrag{U+}{$\avg{u}^+$}
\psfrag{y+}{$y^+$}
\includegraphics[width=1\textwidth]{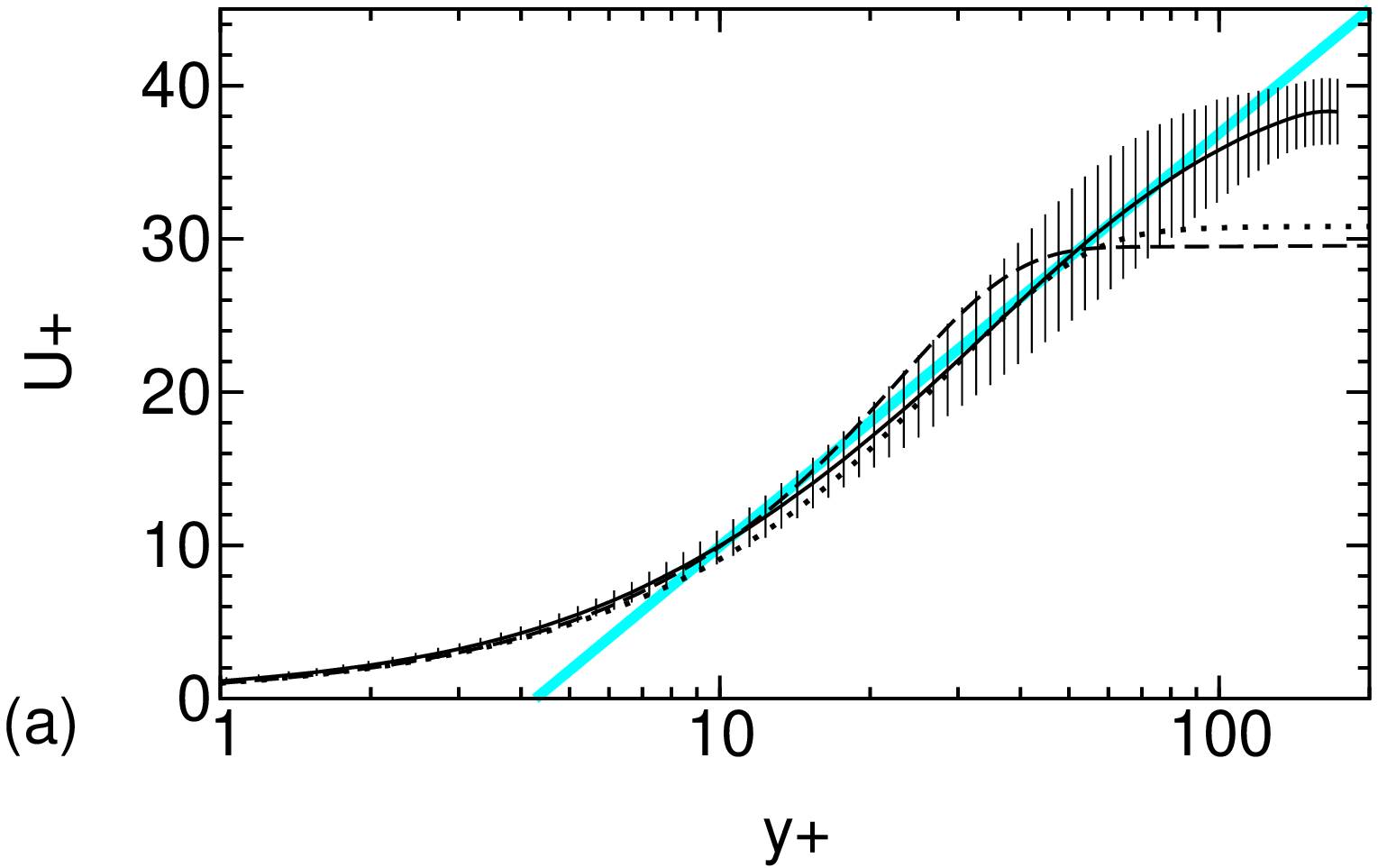}}
\end{minipage}
\hfill
\begin{minipage}{0.32\textwidth}
\subfloat{\label{fig:trans-post--uv+}
\psfrag{(a)}{(b)}
\psfrag{-uv+}{$-\avg{uv}^+$}
\psfrag{y+}{$y^+$}
\includegraphics[width=1\textwidth]{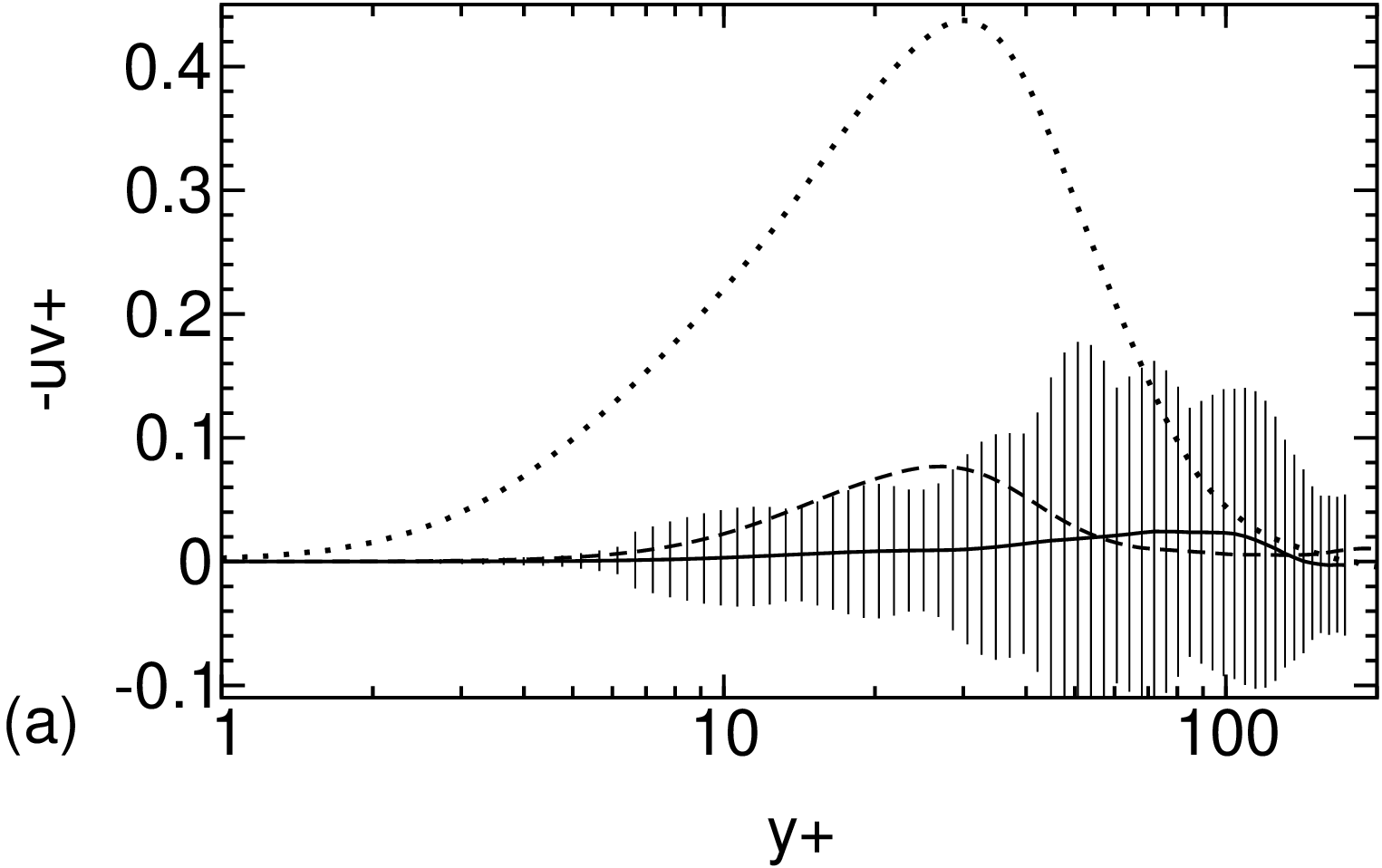}}
\end{minipage}
\hfill
\begin{minipage}{0.32\textwidth}
\subfloat{
\psfrag{(a)}{(c)}
\psfrag{-(dP/dx)x103}{$-(dP/dx)\times10^3$}
\psfrag{tUb/h}{$tU_b/h$}
\psfrag{y}{$y^+$}
\includegraphics[width=1\textwidth]{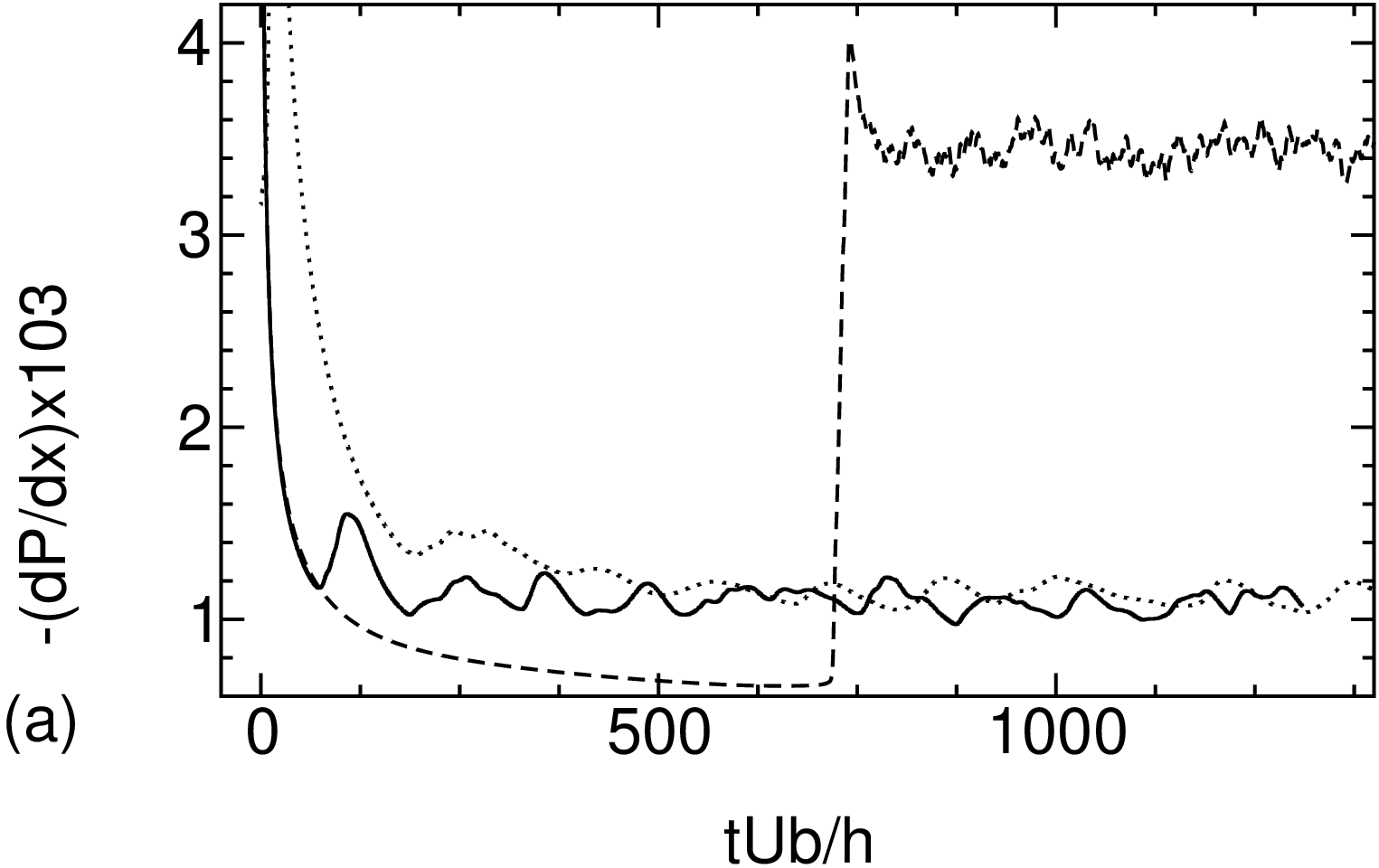}}
\end{minipage}
  \end{center}
    \caption{\label{fig:DR}Polymeric flows at MDR. (a) Shows MDR flow time and space-averaged mean velocity profile \solid and fluctuations of time-dependent, space-averaged mean velocity (error bars),  \eqnref{eq:virk} $\color{cyan}\boldsolid$\color{black}, $Re_\theta=200$ \dashed, and  $Re_\theta=300$  \dotted. (b) Compares Reynolds shear stress, for the three flow states as shown in (a). (c) Plots the time-evolution of pressure gradient for different flow and initial conditions: \dashed Newtonian initiated with isotropic turbulent entrance flow (turbulent intensity $u'/U_b=0.01$), \solid MDR flow ($L=200,We_\tau=720$) initiated with viscoelastic isotropic turbulent entrance flow ($u'/U_b=0.005$),  and \dotted MDR flow ($L=160,We_\tau=200$) initiated from a fully developed turbulent channel flow with $\ten{C}=\ten{I}$ condition for the polymer field.
    }
\end{figure*}

Channel flow simulations are performed in a cartesian domain defined
by the vector base $(\vex{e}_x,\vex{e}_y,\vex{e}_z)$ where $x$, $y$
and $z$ are the streamwise, wall-normal and spanwise directions,
respectively. The components of the velocity vector $\vex{u}$ are
$u$, $v$, and $w$. For a polymer solution, the flow transport
equations are the  conservation of mass, $\vnabla\cdot\vex{u}=0$,
and transport of momentum:
\begin{equation}
\pdert{\vex{u}}+(\vex{u}\cdot\vnabla)\vex{u}=-\vnabla p+\frac{\beta}{Re}\nabla^2\vex{u}+\frac{1-\beta}{Re}\vnabla\cdot\tentau+g(t)\vex{e}_x, \label{eq:mom}
\end{equation}
where $g(t)=-(dP/dx)$ is used to maintain constant mass flux.
The parameter $\beta$ is the ratio of solvent viscosity to the zero-shear viscosity of the polymer solution and affects both the viscous stress and polymer stress terms in \eqnref{eq:mom}. The polymer stress tensor $\ten{T}$ is computed using the FENE-P (Finite Elastic Non-linear Extensibility-Peterlin) model\cite[][]{bird1987dynamics}:
\begin{equation}
\ten{T}=\frac{1}{We}\left(\frac{\ten{C}}{1-\text{tr}(\ten{C})/L^2}-\ten{I}\right)\label{eq:taup}\;,
\end{equation}
where the tensor $\ten{C}$ is the local conformation tensor of the polymer solution and $\ten{I}$ is the unit tensor. The properties of the polymer solution are $\beta$, the Weissenberg number based on the solution relaxation time $\lambda$ and  integral scales, $We=\lambda U_c/H$, or on viscous scales, $We_\tau=\lambda u_\tau^2/\nu$, and the maximum polymer extension $L$. The FENE-P model assumes that polymers may be represented by a pair of beads connected by a non-linear spring and defined by the end-to-end vector $\vex{q}$. The conformation tensor is the phase-average of the tensorial product of the end-to-end vector $\vex{q}$ with itself, $\ten{C}=\langle\vex{q}\otimes\vex{q}\rangle$ whose transport equation is
\begin{equation}
\pdert{\ten{C}}+(\vex{u}\cdot\vnabla)\ten{C}=\ten{C}(\vnabla\,\vex{u})+(\vnabla\,\vex{u})^\text{T}\ten{C}-\tentau\;. \label{eq:C}
\end{equation}
On the left hand side of \eqnref{eq:C}, the first two terms are
responsible for the stretching of polymers by hydrodynamic forces,
whereas the third term models the internal energy that tends to
bring stretch polymers to their least energetic state (coiled). The
FENE-P model has demonstrated its ability to capture the physics of
polymer drag reduction
\cite{sureshkumar1997direct,de2002dns,ptasinski2003tcf,min2003mdr,dubief2004cdr}.
Eqs.~(\ref{eq:mom}-\ref{eq:C}) are solved using finite differences
on a staggered grid and a semi-implicit time advancement scheme described elsewhere\cite{dubief2005nai}.

Here, we build on previous work\cite{dubief2004cdr,dubief2005nai} in
a channel flow at $Re_\tau=hu_\tau/\nu=300$ simulated in a
computational domain of dimensions $(20/3) h\times2h\times(10/3)
h$ or $2000\times 600\times 1000$ in Newtonian wall units, and with periodic conditions in $x$ and $z$. The bulk Reynolds number $Re_b=U_bh/\nu$ is 5000 where $U_b$ and $h$ are the bulk velocity and channel half-height, respectively. Production runs
were performed in a $256\times 161\times 256$ grid.


The initial flow and polymer conditions consist of a constant-mass flow, three-dimensional isotropic turbulence generated by Kolmogorov forcing\cite{boffetta2005drag}. The computational domain and grid are those of the channel with periodic conditions in the wall-normal direction. At time $t=0$, the wall-normal boundary conditions are switched to no-slip to mimic the entrance of a channel flow, i.e. the growth and merger of boundary layer, and a bypass transition induced by free stream turbulence similar to Wu \& Moin\cite{wu2009direct}. The turbulent intensity of the initial condition is 0.5\% for the polymer solution of interest.

A large parameter space was investigated with $L\in[100,200]$ and
$We_\tau\in[120,720]$. All simulations for $We_\tau\geq120$ and
$L\geq160$ or $We_\tau\geq300$ and $L\geq100$ lead to drag reduction in
the range 68 to 72\% and mean velocity profiles in close agreement
with Virk's asymptotic profile. This extensive study will be
discussed in future publications. This
letter focuses on one simulation with $L=200$ and $We_\tau=720$, whose Reynolds shear stress is similar to simulations for $We_\tau \geq 200$.

The mean velocity and Reynolds shear stress of our polymeric
simulation were averaged over homogenous directions and a time duration
of 1500$h/U_b$ of the MDR flow at steady-state. In order to evaluate the envelope of MDR turbulence (i.e., the extent over which the flow oscillates around MDR), we compute both
time- and space-averaged quantities (denoted by $\avg{\bullet}$) and time-dependent quantities averaged  over homogenous directions (denoted by $\spavg{\bullet}$).  The latter statistics are calculated over 300 successive snapshots, separated by $5h/U_b$. The largest positive and negative deviations of $\spavg{u}$ and $-\spavg{uv}$  from $\avg{u}$ and $-\avg{uv}$ are displayed in \figref{fig:DR} by
error bars. \figref{fig:DR}a compares our simulation's mean velocity
profile to \eqnref{eq:virk} and the Newtonian transition data of
\figref{fig:Wu-MDR}. The agreement with Virk is excellent and the
error bars span across Wu \& Moin's $Re_\theta=200$ to 300 states.
The Reynolds shear stress $-\avg{uv}$ in \figref{fig:DR}b
is essentially negligible, yet the largest deviations of $-\spavg{uv}$ exhibit magnitude slightly larger than the $Re_\theta=200$
Newtonian transitional state. The largest negative and positive deviations of the mean velocity profiles are, not surprisingly, found to correlate with lower and higher values of $-dP/dx$ or drag, whose time evolution is shown in \figref{fig:DR}c. Conversely, the largest positive deviations of $-\spavg{uv}$ are correlated to higher drag event. \figref{fig:DR}c shows that, for both initial conditions envisioned in our two thought experiments (fully developed turbulent flow and entrance flow), the flow oscillates between these lower and higher drag events around MDR, as anticipated by our thought experiments. Interestingly, the polymeric transition breaks down earlier than Newtonian flows with higher initial turbulent intensities, which is indicative of a possible destabilizing polymeric effects on (pre-breakdown) streaks\cite{dubief2004cdr,dubief2010polymer}. 

\begin{figure}[t]
  \begin{center}
  \psfrag{dU/dz|wall}{$S(t)=\spavg{du/dy_\text{wall}}$}
   \psfrag{NQ0.1}{$N_{Q>0.1}(\%)$}
\includegraphics[width=0.4\textwidth]{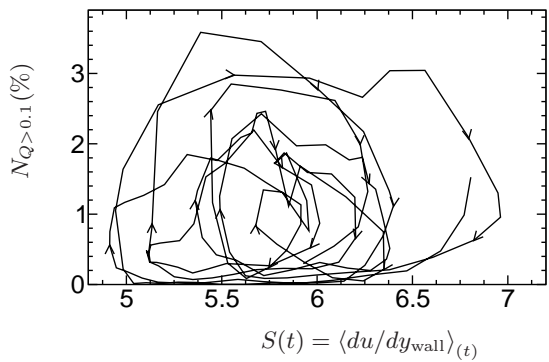}
  \end{center}
    \caption{\label{fig:Q}Orbits of spatially-averaged wall shear in the lower wall of the channel as a function of the volume fraction of the flow region in lower half of the channel occupied by vortices identified by positive regions of the second invariant, $Q>0.1$, of the velocity gradient\cite{dubief2000cvi}.
    }
\end{figure}

Another illustration of the oscillatory nature of MDR turbulence is provided in \figref{fig:Q}, which plots the volume fraction $N_Q$ occupied by vortices (identified using the $Q$ criterion\cite{dubief2000cvi}) in the lower half of the channel as a function of the spatially averaged wall shear $S(t)=\spavg{du/dy_\text{wall}}$. Owing to the inherent subjectivity of threshold selection in any other vortex identification technique\cite{dubief2000cvi}, we tested several threshold in the range $Q\in[0.01,1]$ and verified that (a) all thresholds gave orbits consistent with \figref{fig:Q}  and (b) the chosen threshold $Q=0.1$ depicts regions where velocity vectors exhibit a rotational pattern. The temporal evolution of $N_{Q>0.1}=f(S(t))$ shows clockwise-rotating orbits centered roughly around the time-averaged wall shear $\avg{S}=5.7$. For a typical orbit, starting from a vortex-free state, (i.e., $N_Q=0$), the wall shear decreases to a local minimum, lower that the mean wall shear, at which point vortices appear and grow (i.e., $N_Q>0$). As vortices grow, the wall shear increases. Eventually, vortices are dampened, shown by the reduction of $N_Q$. The wall shear continues to increase toward a local maximum until $N_Q\sim 1\%$, and then decreases back to a local minimum as the vortical activity disappears.

In conclusion, this letter uniquely defines polymer maximum drag
reduction as the transitional state that  corresponds to the onset of the nonlinear
breakdown stage of transition. Furthermore, MDR turbulence corresponds to oscillations between pre- and post-breakdown stages.
These results not only provide a precise definition of MDR they also explain observed differences between experimental data sets at MDR. This letter also achieves a critical step
in the progress toward developing predictive models of polymer drag reduction, as
it narrows the research focus on intermittence and transitional
flows.
 
\begin{acknowledgments}
YD gratefully acknowledges computational support from the
Vermont Advanced Computing Center (supported by NASA award-NNX 06AC88G),  Prof. P. Moin and CTR for their hospitality during the 2010 summer program, and  Prof. X. Wu for sharing his database of transitional boundary layer
flow. 
\end{acknowledgments}

\bibliography{polymers}

\end{document}